\documentclass{PoS}

\usepackage{epsfig,graphicx}
\usepackage{amsmath,amsfonts}

\title{Wilson Fermions, Random Matrix Theory and the Aoki Phase}

\ShortTitle{Wilson Fermions and  Random Matrix Theory}

\author{Gernot Akemann\\
       Department of Mathematical Sciences \& BURSt Research Center\\
Brunel University West London, Uxbridge UB8 3PH, United Kingdom\\
       E-mail: \email{gernot.akemann@brunel.ac.uk}}
\author{{Poul H. Damgaard}\\
       The Niels Bohr International Academy and Discovery Center\\
       The Niels Bohr Institute,
       Blegdamsvej 17, DK-2100 Copenhagen, Denmark\\ 
       E-mail: \email{phdamg@nbi.dk}}
\author{Kim Splittorff\\
       The Niels Bohr Institute,
       Blegdamsvej 17, DK-2100 Copenhagen, Denmark\\
       E-mail: \email{split@nbi.dk}}
\author{\speaker{Jacobus Verbaarschot}\\
       State University of New York\\
       Department of Physics and Astronomy\\
       Stony Brook, NY 11794-3800, USA\\
       E-mail: \email{jacobus.verbaarschot@stonybrook.edu}}

\abstract{
The QCD partition function for the Wilson Dirac operator, $D_W$, 
at nonzero lattice spacing $a$ can be expressed
in terms of a chiral Lagrangian as a systematic expansion in the
quark mass, the momentum and $a^2$. Starting from this chiral Lagrangian 
we obtain an analytical expression  for the spectral density 
of $\gamma_5 (D_W+m)$ in the microscopic domain. 
It is shown that the $\gamma_5$-Hermiticity of the Dirac operator
necessarily leads to a coefficient of the $a^2$ term that is consistent
with the existence of an Aoki phase. The transition to the Aoki phase
is explained,  and the interplay of the index of $D_W$ 
and nonzero $a$ is discussed. 
We formulate a random matrix theory for the Wilson Dirac operator
with index $\nu$ (which, in the continuum limit, becomes equal to the
topological charge of gauge field configurations). It is shown by
an explicit calculation that this random matrix theory reproduces the 
$a^2$-dependence of the chiral Lagrangian in the microscopic domain, and 
that the sign of the $a^2$-term is directly related to the  
$\gamma_5$-Hermiticity of $D_W$. 

}

\FullConference{The XXVIII International Symposium on Lattice Field Theory, Lattice2010\\
		June 14-19, 2010\\
		Villasimius, Italy}

\begin{document}

\newcommand{\be}{\begin{eqnarray}}
\newcommand{\ee}{\end{eqnarray}}
\newcommand\del{\partial}
\newcommand\nn{\nonumber}
\newcommand{\Tr}{{\rm Tr}}
\newcommand{\Str}{{\rm STr}}
\newcommand{\mat}{\left ( \begin{array}{cc}}
\newcommand{\emat}{\end{array} \right )}
\newcommand{\vect}{\left ( \begin{array}{c}}
\newcommand{\evect}{\end{array} \right )}
\newcommand{\tr}{{\rm Tr}}
\newcommand{\hm}{\hat m}
\newcommand{\ha}{\hat a}
\newcommand{\hz}{\hat z}
\newcommand{\hx}{\hat x}
\newcommand{\tm}{\tilde{m}}
\newcommand{\ta}{\tilde{a}}
\newcommand{\tz}{\tilde{z}}
\newcommand{\tx}{\tilde{x}}
 \newcommand{\bitem}{\begin{itemize}}
\newcommand{\eitem}{\end{itemize}}
 \newcommand{\bmini}{\begin{minipage}}
\newcommand{\emini}{\end{minipage}}
 \newcommand{\dl}{$}
\newcommand{\dr}{$}


\section{Introduction}

For sufficiently small eigenvalues 
the spectrum of the QCD Dirac operator can be obtained 
from the zero momentum part of a chiral Lagrangian that
is consistent with the global symmetries of QCD \cite{SV}. The reason
is spontaneous symmetry breaking and the existence of a mass gap,
so that QCD at low energy is described by a chiral Lagrangian
for the Goldstone bosons.
If the Goldstone bosons corresponding to the mass scale of the Dirac
spectrum we are interested in, have a Compton wavelength that is much large
than the size of the box, the zero momentum part of the partition function, 
which determines its
mass dependence, 
factorizes from
the nonzero momentum part \cite{GL}.
All theories with a mass gap and the global symmetries 
and symmetry breaking pattern of QCD are described
by the same zero momentum chiral Lagrangian. Analyzing the simplest
such theory, chiral random matrix theory, allows us to 
answer questions that cannot be addressed otherwise.   

The Dirac operator in lattice QCD has discretization errors, and if
the discretization errors change the underlying symmetries, the low lying Dirac
spectrum will be affected significantly. This is the case for the Wilson Dirac 
operator
$D = D_W + m$, where
\be
D_W = \frac{1}{2}\gamma_{\mu}(\nabla_{\mu} + \nabla^*_{\mu})
-\frac{1}{2}a\nabla^*_{\mu}\nabla_{\mu}
\ee
 is written in terms
of forward ($\nabla_{\mu}$) and backward ($\nabla^*_{\mu}$) covariant 
derivatives,
and $m$ is the quark mass. 
The Wilson term violates the axial symmetry but the Dirac operator
remains $\gamma_5$ Hermitian
\be
D_W^\dagger = \gamma_5 D_W \gamma_5.
\ee
Because of this, the operator $D_5 \equiv \gamma_5(D_W+m)$ is Hermitian,
and is computationally much simpler to work with. Its spectrum
has been analyzed in lattice QCD simulations \cite{Luscher,Luscher:2007se} and 
by means of mean  field studies of chiral Lagrangians \cite{Sharpe2006,Golterman}.

\begin{center}
\begin{figure}[h!]
{\includegraphics[width=4.5cm,clip]{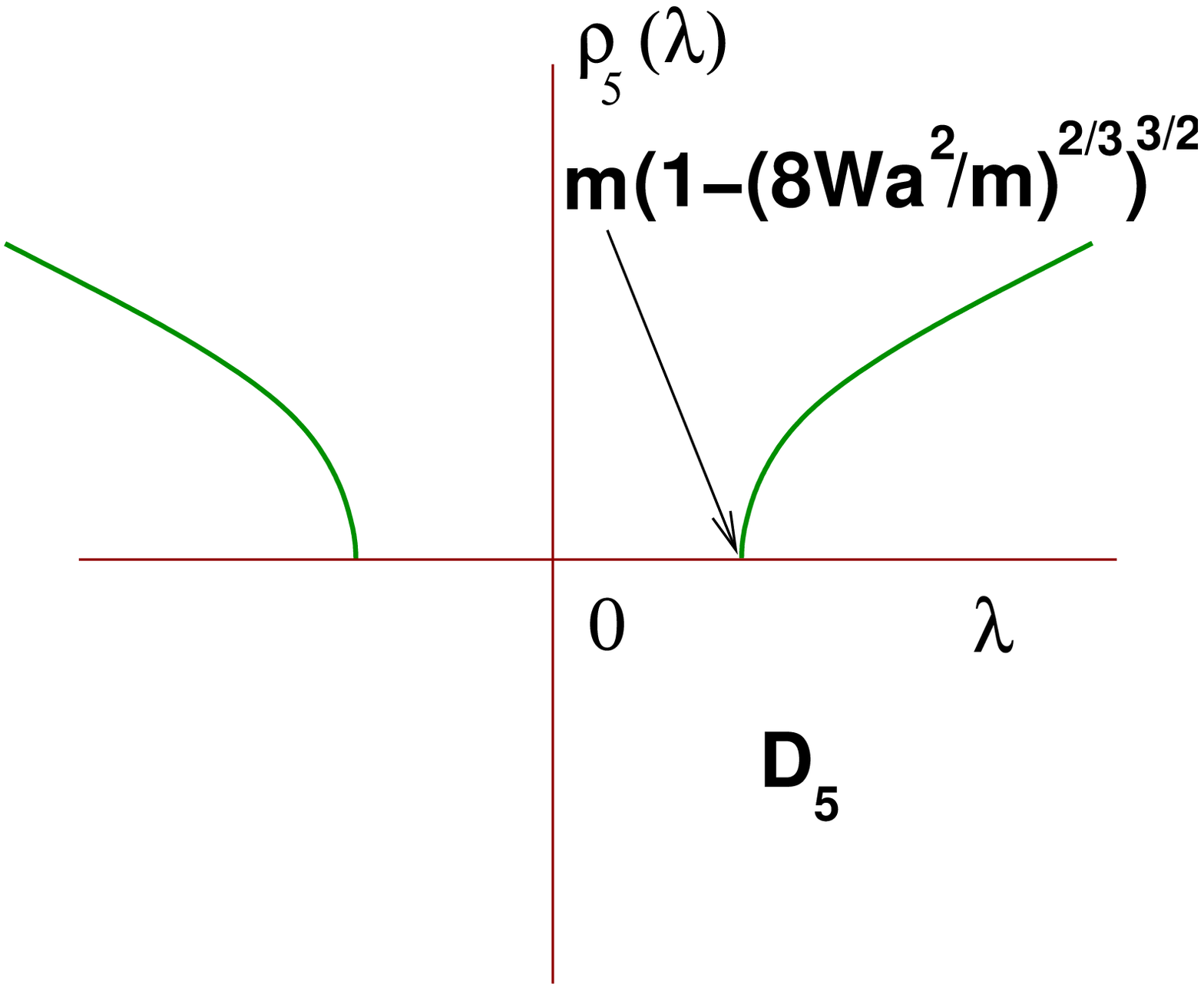}}\hspace*{0.5cm}
{\includegraphics[width=4.5cm,clip]{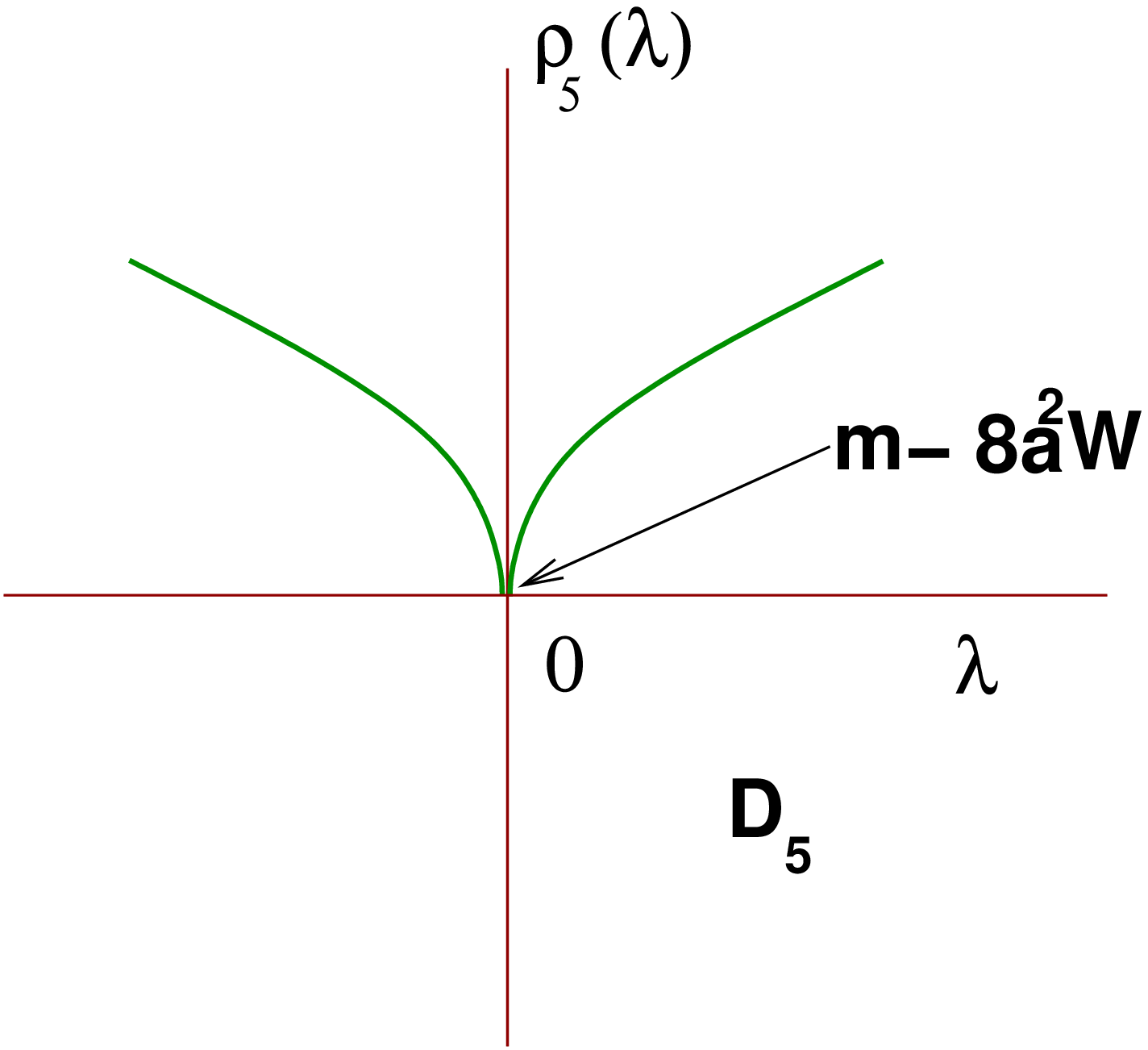}}\hspace*{2cm}
{\includegraphics[width=2.5cm,clip]{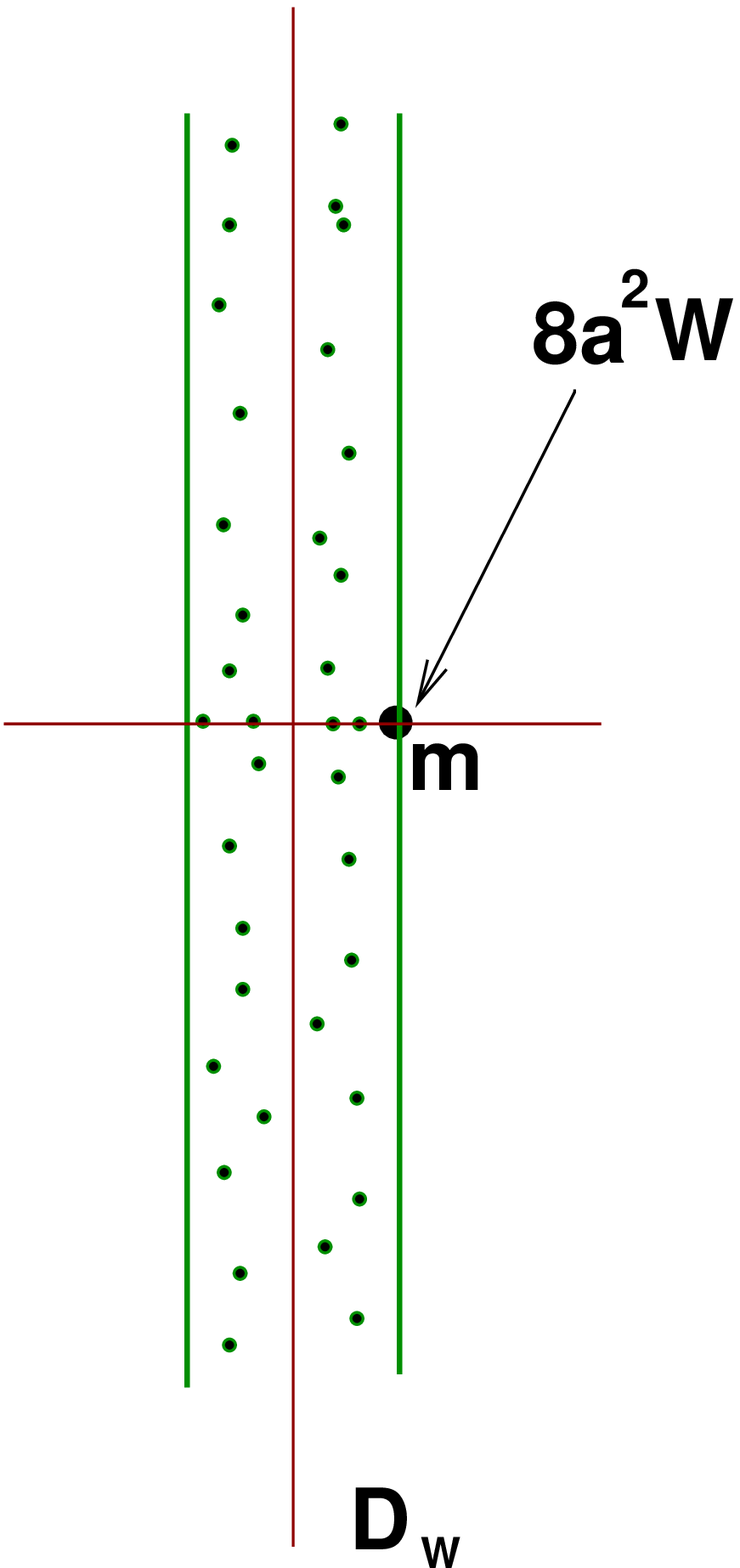}}
\caption{Schematic picture of mean field results for
the spectral density of $D_5$ (left two figures) and
the spectrum of $D_W$ (left). 
The right two figures show the onset of the
Aoki phase. In this Figure, $W= W_8/\Sigma$.}
\label{fig:fig1}
\end{figure}
\end{center}

For $a=0$, the spectrum of $D_5$ has a gap $[-m, m]$. At nonzero $a$,
states intrude inside the gap, and for sufficiency large $a$, the gap closes,
and we enter what is known as the Aoki phase (see Fig. \ref{fig:fig1}). In terms
of $D_W$, this is the point where $m$ hits the cloud of eigenvalues (see  Fig. \ref{fig:fig1}, right).
The sharp edge of the spectrum of $D_5$  in the mean field limit
smoothes out for finite $V$. In lattice QCD simulations \cite{Luscher}, 
a roughly Gaussian tail inside the gap with a width that scales
as $ a/\sqrt V$ is observed. We will provide an analytical explanation of this result
by means of a chiral Lagrangian for the Wilson Dirac spectrum. 
We will also construct a random matrix theory
 with the symmetries of the Wilson Dirac operator at $a \ne 0 $
that reproduces the zero momentum part of the chiral Lagrangian for the Wilson 
Dirac operator
including $O(a^2)$ terms.

\section{Chiral Lagrangian}
The chiral Lagrangian for Wilson fermions is an expansion in the mass, 
the momentum and the lattice spacing. Up to a few low-energy constants
it is determined uniquely by  the transformation
properties of the QCD partition function
\cite{SharpeSingleton,RS,BRS}.
We decompose the  partition function as \cite{DSV}
\be
Z(m,z; a)  = \sum_\nu Z_{N_f}^\nu(m,z;a).
\ee
In the microscopic domain, where $mV$, $zV$ and $a^2V$ are kept
fixed for $V \to \infty$, the partition function with index $\nu$ is
given by (we will see below that $\nu$ is the index of the Dirac operator)
\be 
Z_{N_f}^{\chi,\,\nu}(m,z;a) & = & \int_{U(N_f)} d U \ {\det}^\nu U
 \ e^{V [\frac 12 (m+z) \Sigma { \rm Tr} U
+\frac 12 (m-z)\Sigma{ \rm Tr} U^\dagger 
    - a^2 W_8{\rm   Tr}(U^2+{U^\dagger}^2)
]}.
\label{Zfull}
\ee
Here, $\Sigma$ is the chiral condensate, $W_8$ is a low-energy constant and
$z$ is the axial mass.
Terms that can be expressed as squares of traces of $U$ and $U^\dagger$
are possible \cite{ADSV}, but they are suppressed in  the the large $N_c$-limit 
\cite{KL} and are not considered in this talk.
It is instructive to perform a mean field analysis of this chiral Lagrangian
valid in the thermodynamic limit at fixed $m$ and $a$. It turns out that the  
properties of the partition function
depend crucially on the sign of $W_8$. 

\noindent
$W_8 >0$: For $z = 0$ the partition function shows a second order
phase transition at $m\Sigma=8a^2 W_8$. At this point the quark mass
hits the cloud of eigenvalues of $D_W$.
The spectrum of $D_5$ follows from the $z$-dependence of the partition
function. It has a gap $[-z_g, z_g]$ that closes  at the critical point,
where
\be 
z_g = m(1- (8a^2 W_8/m\Sigma)^{2/3})^{3/2}.
\ee

\noindent
$W_8<0$: For $z=0 $ the partition function does not undergo a phase transition. This implies that the eigenvalues of $D_W$ are not scattered in the
complex plane which is consistent with an anti-Hermitian Dirac operator, and
violates the very premise we started form. 
One possibility is that $D_W$ would be  both anti-Hermitian and 
$\gamma_5$-Hermitian, but then $\{\gamma_5,D_W\}= 0$, which is not the case.

\subsection{QCD inequalities and the sign of $W_8$}
 
A direct consequence  of $ D_5^\dagger = D_5 $ is
the QCD inequality
\be
Z_{N_f=2}^{{\rm QCD}, \nu}(m,z) =\langle {\det}^2(\gamma_5(D_W +m)+z)\rangle > 0 \qquad {\rm for} \quad m, z \quad {\rm real}.
\ee
For an anti-Hermitian Dirac operator we have that
\be
i^{-2\,{\rm dim}(D)}Z_{N_f =2}^{{\rm QCD}, \nu}(m,z) =i^{-2\,{\rm dim}(D)}\langle {\det}^2((D_W +im)+iz)\rangle > 0 \qquad {\rm for} \quad m, z \quad {\rm real},
\ee
where ${\rm dim}(D)$ is the total dimension of the Dirac matrix. In the continuum limit,
$i^{2\,{\rm dim}(D)} = (-1)^\nu$.

\begin{center}
\begin{figure}[h!]
\includegraphics[width=5.5cm,angle=-90,clip]{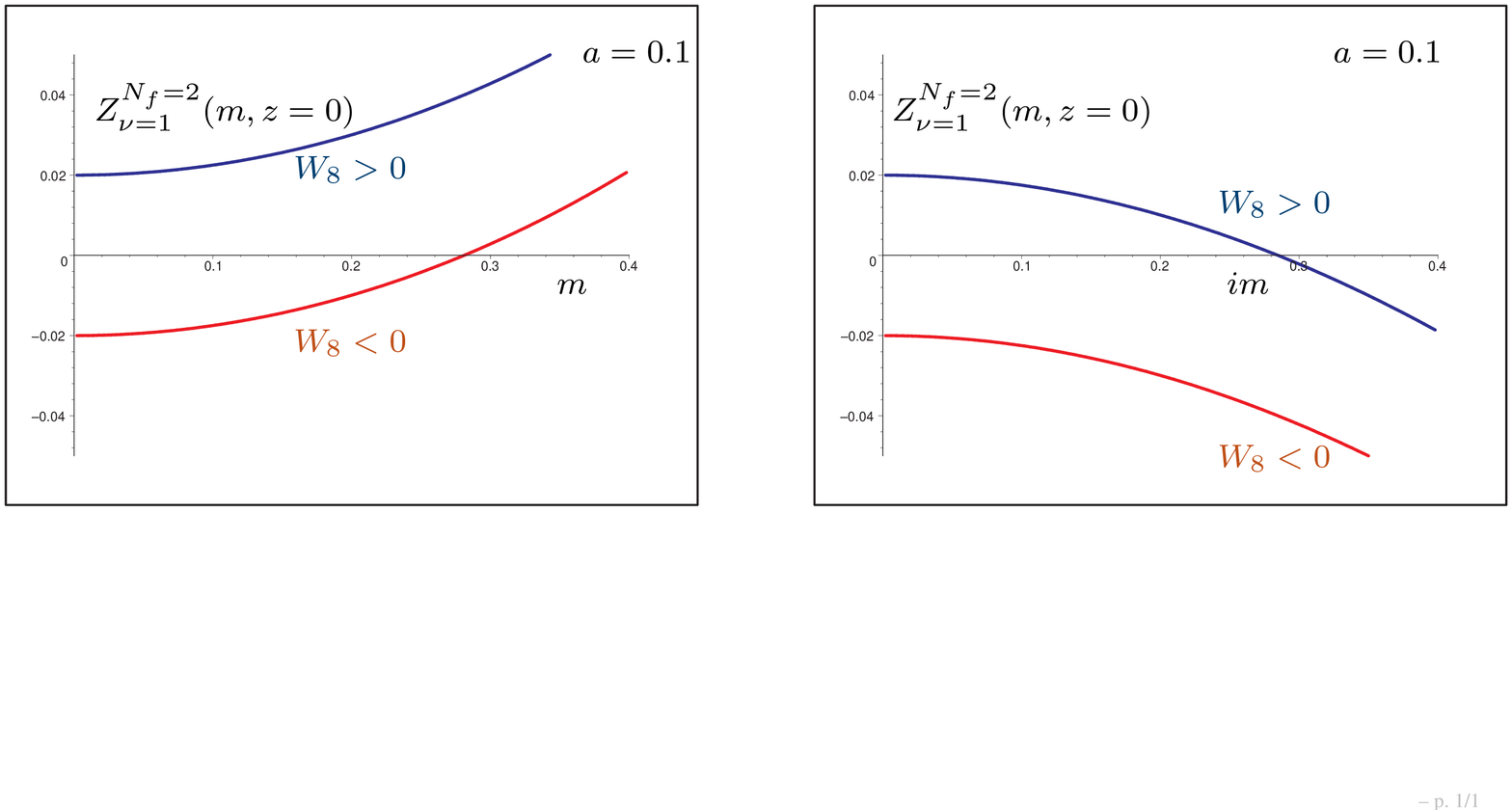}\\[-1.0cm]
\caption{
 The mass dependence of the partition function (2.2) 
for
real mass  and imaginary mass.}
\label{fig:fig2}
\end{figure}
\end{center}

By changing variables $U \to iU $ in the chiral partition function (\ref{Zfull}) 
it follows that
\be
Z^{\chi, \,N_f}_\nu(0,0, W_8) = (i)^{N_f \nu}Z^{\chi,\, N_f}_\nu(0,0, -W_8).\nn
\ee
Since for large mass the partition functions for $+W_8$ and $-W_8$ have the
same sign, one of them must change sign as a function of $m$, and not both signs
of $W_8$ can be allowed by the QCD inequalities.
In Fig.
\ref{fig:fig2} we show the $N_f=2$ partition function for $\nu = 1$
as a function of $m$ (left) and as a function of $im$ (right). The blue
curves show the $W_8>0$ results, whereas  $W_8< 0$ for the red curves.
 \vspace*{-1.0cm}
 
The conclusion is that $W_8>0$ can only correspond to a $\gamma_5$-Hermitian
Dirac operator, whereas for $W_8<0$ the chiral partition function only has a
definite sign for imaginary mass so that the corresponding Dirac operator
must be anti-Hermitian.

\section{Dirac Spectrum}
To find the Dirac spectrum of $D_5$ at $z$ we have to add a fermionic quark with axial mass $z$
and
a bosonic quark with mass $z'$ to the partition function so that the generating
function is given by
\be
Z^{{\rm QCD},\nu}(m,m',z,z'; a) =\left \langle\frac{\det(\gamma_5(D_W+m) +z)}
{\det(\gamma_5(D_W+m') +z'-i\epsilon)} 
\prod_f \det(D_W+m_f)
\right \rangle.
\label{zpart}
\ee
The infinitesimal $i\epsilon$ term is essential. The resolvent
\be
\left . - \frac d {dz'} Z^ {{\rm QCD},\nu}(m,m,z,z';a) \right |_{z'=z} = 
 \left \langle {\rm Tr} \frac 1{\gamma_5(D_W+m) +z-i\epsilon} 
\prod_f \det(D_W+m_f)
\right \rangle.
\label{zgen}
\ee
is discontinuous across  the support of the spectrum, and for real $z$ its imaginary
part gives the spectral density of the Dirac operator 
(denoted by $\rho_5^\nu(z)$).
The chiral Lagrangian corresponding to the generating function (\ref{zpart})
is uniquely determined by symmetries. What is not determined by symmetries is the integration contour of the group manifold. For compact integrals this is not
an issue, but for noncompact integrals corresponding to bosonic quarks, a change in
integration contour may lead to a different analytical continuation of the integral
in its complex parameter plane \cite{DOTV,SV-fac,Golterman}. 
When $D_W$ is Hermitian or anti-Hermitian it is possible to analytically continue
the partition function in the entire complex $m$-plane. When $D_W$ is non-Hermitian 
the generating function is not analytic if $m$ is
inside the support of the spectrum and the expression (\ref{zgen}) cannot be used to derive the Dirac spectrum.  

For a  $\gamma_5$-Hermitian  Dirac
operator the partition function (\ref{zpart}) is an analytical function of $z$, and
only the part of the complex $a$-plane
to which the partition function can be continued analytically can correspond to
lattice QCD at $ a\ne 0$. In particular, analyticity in $z$ requires that 
${\rm Re}(a^2W_8) >0$.

The second interesting observable that can be  calculated from the generating function is
\be
\Sigma^\nu(m) = \left. \frac d{dm} \log Z^\nu(m,m',z,z';a) \right |_{z'=z=0, m'=m}.
\ee
For small $a$ it can be shown that the discontinuity of 
this quantity accross the real axis gives the distribution of the chirality over the real eigenvalues of $D_W$,
\be
  \rho^\nu_\chi(\lambda^W) = \left \langle \sum_{k, \lambda_k^W 
 \in {\mathbb R}} \delta(\lambda^W-\lambda_k^W)
{\rm sign}(\langle k| \gamma_5 | k \rangle) \right \rangle,
\label{index}
\ee
where $\lambda_k^W$ are the eigenvalues of $D_W$.
The integral $\int d\lambda^W  \rho_\chi^\nu (\lambda^W) $  is equal to the 
index of the Dirac operator. It is 
defined configuration by configuration as
 $  \sum_{k, \lambda_k^W  \in {\mathbb R}} 
{\rm sign}(\langle k| \gamma_5 | k \rangle) $.

In the microscopic domain of QCD, the generating function
(\ref{zgen})
is given by an integral over the zero momentum part of the Goldstone manifold
\be
\label{ZSUSY}
Z^{{\rm SUSY}, \, \nu}_{1|1}(m,m',z,z';\ha)  & = & \int  dU \
{\rm Sdet}(U)^\nu 
  e^{i\frac{1}{2}{\Str}({\cal M}[U-U^{-1}])
    +i\frac{1}{2}{\Str}({\cal Z}[U+U^{-1}])
    +\ha^2{\Str}(U^2+U^{-2})}. 
\ee
 The integration is
over the maximum Riemannian graded submanifold of
$Gl(1|1)$ \cite{DOTV} and
can be worked out analytically. The result is a sum of  products of simple one-dimensional 
integrals \cite{DSV}. Calculating the integral $\int d\lambda^W 
\rho_\chi^\nu(\lambda^W)$
we find that it is identically equal to $\nu$. This shows that the
parameter $\nu$ in the chiral Lagrangian is the index of the Dirac
operator defined below Eq. (\ref{index}).

It is straightforward to extend this result for QCD with dynamical quarks. 
The result for $N_f = 1$ is discussed in \cite{ADSV-2} in this volume.

\section{Random Matrix Theory}
Since the chiral Lagrangians (\ref{Zfull}) and (\ref{ZSUSY}) are
 determined uniquely by the global symmetries of QCD, they can also
be obtained from a random matrix theory with the same symmetries.
The random matrix partition with index $\nu$
is defined by \cite{DSV}
\be
{Z}^{{\rm RMT},\, \nu}_{N_f} = \int d{A}d{B}d{W}  \  
{\det}^{N_f}({D}_W + {m} + {z}{\gamma}_5) 
\ P({D}_W),
\label{zwrmt}
\ee
with
\be
D_W = \mat a A & C \\ -C^\dagger & aB \emat  
\qquad
{\rm and} \quad
 A^\dagger = A, \qquad B^\dagger = B . \nn
\ee 
Here,  $ A $ is a square matrix of size $ n \times n $, and $ B$
is a square matrix of size $ (n+\nu) \times (n+\nu)$. The matrix $C$ is
a complex $ n \times (n+\nu)$ matrix. The matrix elements of $D_W$
are taken to be  distributed with Gaussian weight
\be
\label{P}
P(A,B,W) \equiv e^{-\frac {N}{4}{\rm Tr}[A^2+B^2] 
-\frac N2 {\rm Tr} [ W W^\dagger]},
\ee
where $N=2n+\nu$. For $a \to 0$ the random matrix Dirac operator $D_W$ has exactly
$\nu$ zero eigenvalues. For nonzero $a$, the index of the Dirac operator
defined below Eq. (\ref{index}) is exactly equal to
$\nu$ for each matrix in the ensemble.

In the microscopic domain, this random matrix theory partition 
function reduces to the chiral Lagrangian introduced 
in Eq. (\ref{Zfull}) with $\Sigma=N/V$ and $W_8 = \frac 12a^2 N$. Notice
that $W_8$ is positive.

\section{Results for the Dirac Spectrum}

Analytical expressions for $\rho^\nu_5(\lambda_5)$ and $\rho_\chi^\nu(\lambda_W) $ 
were derived from (\ref{ZSUSY}) in \cite{DSV}. Plots of $\rho^\nu_5(\lambda_5)$
are shown in Figs. 3 and 4.
For $a = 0$ the eigenvalue  density of $ D_5 $ can be decomposed as
\be
\rho_5^\nu(\lambda_5) = \nu \delta (\lambda_5- m) + \rho_{\lambda_5> m}(\lambda_5).
\ee
For $a\ne 0$ the width of the peak at $x=m$ becomes finite.
 In Fig. 3
we show the spectral density of $D_5$ for $\nu = 1$ and $ a= 0.05$. Also
shown is the $a^2 W_8 V \ll 1$ analytical  result for the density of the ``topological'' eigenvalues
for $\nu = 1$ 
which is given by a simple Gaussian
\be
\rho^{\nu=1}_{5, {\rm topo}}(\lambda_5) = \frac 1{4a\sqrt{ \pi V W_8}} e^{-\frac {V\Sigma^2 
(\lambda_5-m)^2}{16a^2W_8}}.
\ee
~For small $ a $, it can be shown for arbitrary $\nu$ that  this distribution 
approaches the spectral 
density of the real eigenvalues of $D_W $ up to a shift by $m$ \cite{ADSV}.
This figure suggests that it is possible to extract the 
low-energy constant $ W_8$ from the width of
the distribution of the smallest eigenvalue.

\bmini{6.5cm}\hspace*{-2cm}
\baselineskip 8pt
\includegraphics[height=8cm,angle=-90,clip]{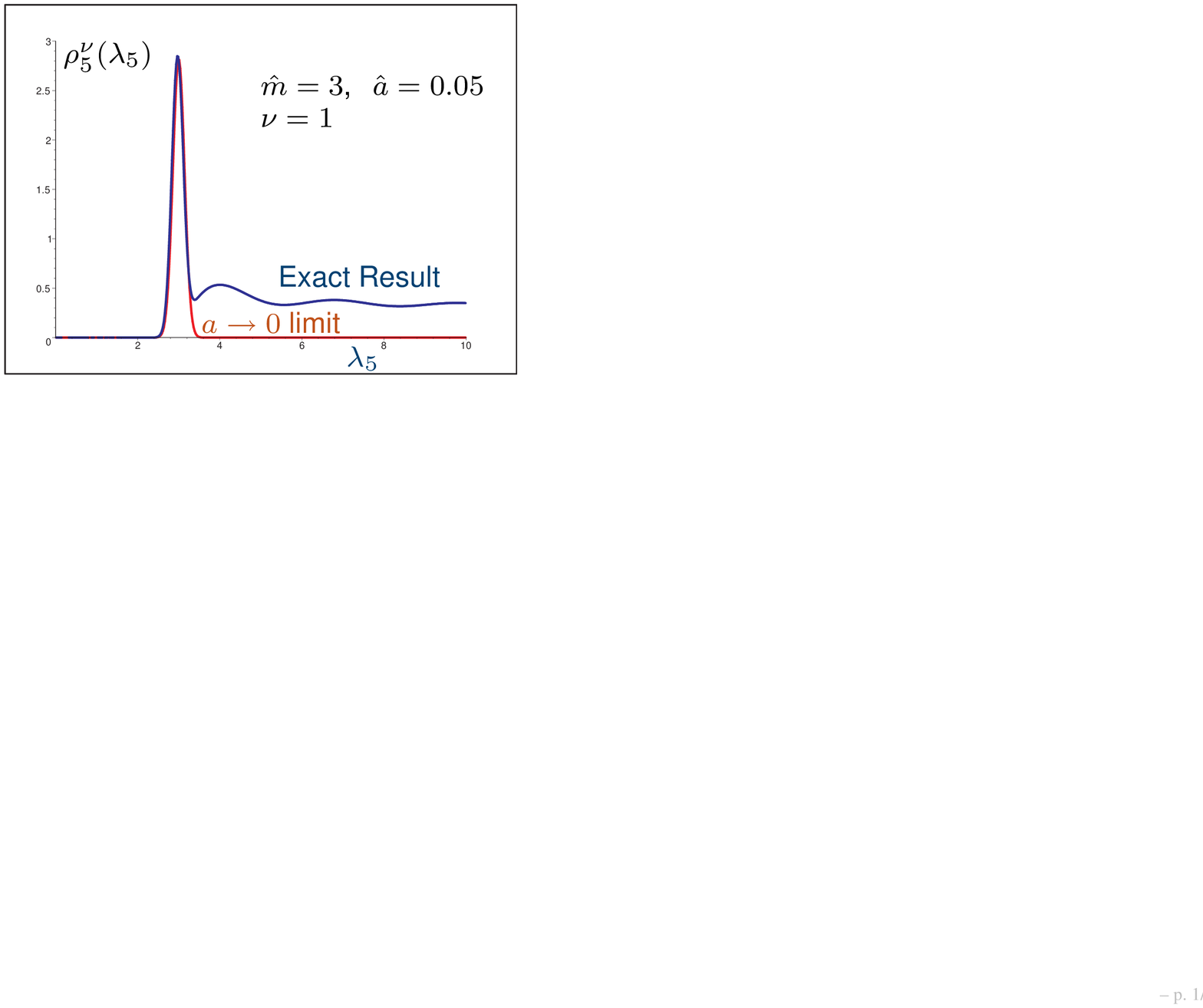}\\[-0.8cm]
{\small {\bf Figure 3:} 
Microscopic spectral density of $\gamma_5(D_W+m)$ for $\nu =1$ (blue curve). Also shown
is the distribution of the smallest eigenvalue for $\nu = 1$ obtained in the
$ a\to 0$ limit.}

\emini
\hspace*{0.5cm}
\bmini{6.5cm}
\baselineskip 8pt
\includegraphics[width=6cm,clip]{wilson-ga5Dw-nu2-3as.eps}\\
{\small {\bf Figure 4:}
The microscopic spectrum of $ \gamma_5(D_W+m)$
    for $ mV\Sigma =3$, $\nu=2$ and  
$a\sqrt{W_8 V}=0.125,\; 0.250$ and $ 0.500$,
  respectively.}
\emini\\[0.5cm]

For  $|\lambda_5-m|/a$  fixed for $a \to 0 $
the spectral density inside the gap
can be obtained from a saddle point analysis:
\be
\rho_5(\lambda_5) \sim e^{-{\Sigma^2 V ( \lambda_5- m)^2}/{16 a^2 W_8}}\qquad 
{\rm for} \quad  0< x \ll  m. \nn
\ee
The width parameter is given by 
$
\sigma^2 =  {8a^2 W_8}/{V \Sigma^2}$, and the ratio of $\sigma$ and
and $\Delta \lambda = \pi/\Sigma V$ is equal to
${\sigma}/{\Delta \lambda}
= {(\sqrt 8/\pi}) a \sqrt{W_8 V}$
\cite{DSV}.
This is exactly the scaling behavior found in lattice QCD simulations 
\cite{Luscher}. We note that in this reference the results are an average over
different values of $\nu$. Our results show that, in particular for small $a$,
the Wilson Dirac spectrum shows a strong dependence on $\nu$, and it would be
very interesting to analyze the lattice results for fixed index of the
Dirac operator.

\section{Conclusions}

Using chiral perturbation theory we have obtained exact analytical expressions
for the spectrum of the Wilson Dirac operator in the microscopic domain. Our result
are obtained for fixed value of the index of the Dirac operator and corrections
up to $O(a^2)$ have been included. We have formulated a random matrix theory that
in the limit of large matrices reproduces the chiral Lagrangian and the analytical
results for the Dirac spectrum. Using QCD inequalities for the Wilson Dirac operator
we find that the sign of the coefficient of $O(a^2)$ in the 
chiral Lagrangian is consistent with the existence of an Aoki phase.
In the limit of small $a$ we have obtained analytical results for the distribution
of the smallest Dirac eigenvalues with fixed nonzero index of the 
Dirac operator.
We find that the width of the distribution scales as $a/\sqrt V$ in agreement with 
earlier lattice simulations. Our results make it possible to extract additional
low-energy constants of Wilson chiral perturbation theory from the distribution of
the smallest Dirac eigenvalues.

{\bf Acknowledgments:}
We would like to thank many participants of the Lattice 2010 Symposium and
the CERN TH-Institute ``Future directions in lattice
gauge theory - LGT10'' for discussions. Two of us (GA and JV) thank 
the Niels Bohr Foundation for support and the Niels
Bohr Institute and the Niels Bohr International Academy for its
hospitality. This work was supported  by U.S. DOE Grant No. 
DE-FG-88ER40388 (JV) and the 
Danish Natural Science Research Council (KS).

\end{document}